# To the electrostrictive mechanism of nanosecond-pulsed breakdown in liquid phase

Yohan Seepersad[1,2], Danil Dobrynin[1*], Mikhail Pekker[1], Mikhail N. Shneider[3] and Alexander Fridman[1,4]

[1] A. J. Drexel Plasma Institute, Drexel University, Camden NJ 08103
[2] Electrical and Computer Engineering Department, Drexel University, Philadelphia PA 19104
[3] Mechanical and Aerospace Engineering Department, Princeton University, Princeton, NJ 08544
[4] Mechanical Engineering and Mechanics Department, Drexel University, Philadelphia PA 19104

**Abstract**
In this study we have studied the initial stage of the nanosecond-pulsed discharge development in liquid phase. Modeling predicts that in the case of fast rising strong nonhomogeneous electric fields in the vicinity of high voltage pin electrode a region saturated with nanoscale non-uniformities may be developed. This phenomenon is attributed to the electrostriction mechanisms and may be used to explain development of breakdown in liquid phase. In this work, schlieren method was used in order to demonstrate formation of negative pressure region in liquids with different dielectric permittivity constants: water, ethanol and ethanol-water mixture. It is shown that this density perturbation, formed at the raising edge of the high voltage pulse, is followed by a generation of a shock wave propagating with the speed of sound away from the electrode, with negative pressure behind it.

## 1. Introduction

Generation of plasmas in liquids has been studied extensively in the past for various applications, e.g. water sterilization [1, 2] and removal of contaminants [1-5], insulation and high power switching [3]. Traditionally, plasma is considered as gas phase phenomenon, and until recently its generation directly in liquid phase has never been observed experimentally [6, 7]. In our recent studies, it was demonstrated, that application of short (nanosecond- and even picosecond-range) high voltage pulses when applied to liquid may result in formation of a discharge which exists directly in liquid: no microscopic bubbles were observed in the region of plasma formation before, nor right after the discharge [6, 7].

In general, plasma generation in liquids is traditionally explained via so-called "bubble" mechanism: gas phase regions provide required space for streamer formation via classical mechanism [1, 5, 8]. Formation of bubbles (or voids) in liquid due to application of relatively long (hundreds of nanoseconds and microsecond-range) high voltage pulses has been studied extensively over the last years (see, for example, [5, 8]). In this case, liquid is shown to undergo phase transition and expansion, resulting in delayed formation of a discharge in these low-density regions.

---

[*] Corresponding author: danil@drexel.edu

Another mechanism, which was introduced in [9] and [7], considers the effect of short nanosecond high electric field pulses in liquid which result in formation of nanoscale "pores", or voids in negative pressure regions due to cavitation. It is shown, that electrostriction phenomenon may provide necessary space for electrons to gain energy and for formation of an initial streamer, which then results in generation of a discharge via leader-type mechanism. It is also proposed that in the case of ultra-short (picosecond) voltage pulses, the discharge may actually develop via direct electron impact ionization without phase transition or even formation of nanoscale voids [10].

In this study, we are focused on experimental observation of a region in which formation of nanosecale pores is predicted. Here, we have used nanosecond high voltage pulses applied to liquid in a discharge cell with pin-to-plane electrode configuration. In order to observe this region, we have used schlieren method with under-breakdown conditions. We also compare experimental results with electrostriction model predictions.

## 2. Experimental setup and electrostriction model

The experiments were conducted using pin-to-plane electrode configuration (similar to [7]), where high voltage pin electrode was made mechanically sharpened 1 mm thick tungsten rod to the radius of 35 μm, and placed 3 mm away from 18 mm diameter copper grounded electrode. The liquid layer between the center of the discharge gap and quartz window was 50 mm. In order to study the effect of dielectric properties of liquid we used two liquids in the experiments: distilled deionized (Type II) water (maximum conductivity 1.0 μS/cm, EMD Chemicals), ethyl alcohol (>99.9%, Decon Labs), and their 1-to-1 volume mixture. This allowed us to study liquids with different dielectric permittivity constants: 80, 25 and 55.

In the experiments we used nanosecond pulses with +11.2 kV pulse amplitude in 50Ω coaxial cable (22.4 kV on the high-voltage electrode tip due to pulse reflection), 10 ns pulse duration, 3 ns rise time and 4 ns fall time (Figure 1, a). Pulse frequency was set to 1 Hz. The generator was made by FID Tech Company on the basis of solid-state switches. A 15 m long coaxial cable RG393/U with a calibrated back current shunt mounted in a middle of it was used for control of applied voltage and power measurements.

The discharge visualization measurements were performed using 4Picos ICCD camera from Stanford Computer Optics and 32 mW laser diode operating at 532 nm (Figure 1, b). The camera had an 18mm diameter multi-alkaline photocathode with a spectral response from 180 to 750 nm. The camera's typical field of view was 750x500 μm, and spectral response was 250 – 750 nm taking into account UV-emission absorption in the water layer. The synchronization system for the experiments was built on the basis of a Tektronix AFG-3252 Arbitrary/Function Generator. The generator has synchro-output and two adjustable channels with a signal rise/fall time of less than 2.5 ns and a typical jitter (RMS) less than 20 ps with a delay time resolution of 10 ps.

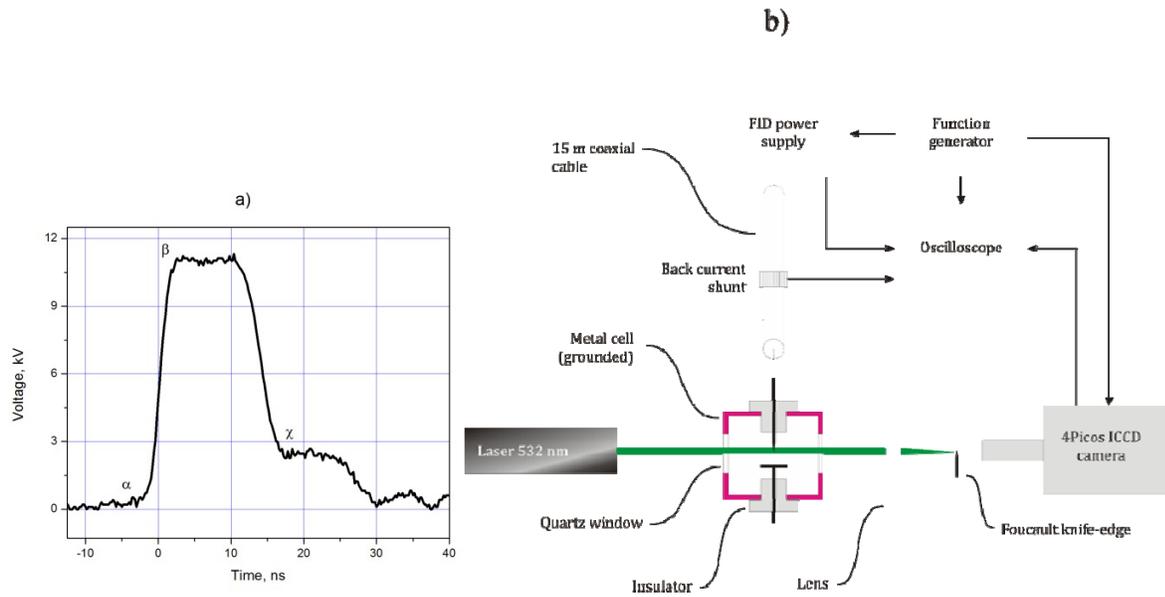

Figure 1

In [9] and [11], it was shown that in the case of fast rise (on the order of a few nanoseconds, points α and χ in Figure 1) of electric field, the fluid lacks time to expand, and the total pressure in the region is negative:

$$P_{total} = P_{hydr} - P_{electr} = P_{hydr} - \rho\varepsilon_0 \left(\frac{\partial\varepsilon}{\partial\rho}\right) E^2, \qquad (1)$$

where $E$ is electric field, $\varepsilon$ – liquid dielectric permittivity constant, and $\rho$ – density of liquid. In order to model the experimental conditions, we have solved the equations of compressible hydrodynamics for dielectric liquid in a pulsed inhomogeneous electric field [9, 11]. As a volumetric force acting on the dielectric fluid a volumetric term in Helmholtz equation related to electrostriction has been taken into account [9, 11-14].

Figure 2 shows pressure distributions in liquid water near the high voltage electrode at the points of maximum electric filed, plateau and trailing edge of the voltage pulse. If the total pressure in liquid water ($\varepsilon = 81$) is less than $-(20 + 30)$ MPa, cavitation processes lead to density perturbations due to electrostriction effect [9, 11]. On the plateau of the high voltage pulse, electric field is constant and hydrodynamic pressure compensates pressure due to electrostriction, so that the total pressure is zero. At the trailing edge of the voltage pulse, total pressure is positive and is equal to the hydrodynamic pressure, leading to formation of a shock wave propagating with the speed of sound with negative pressure behind it [11] (Figure 3). If the absolute value of the negative pressure behind the shock wave is greater that the critical value of ~$(20 + 30)$ MPa, discontinuity of liquid may result in formation of nanoscale pores in this region as well.

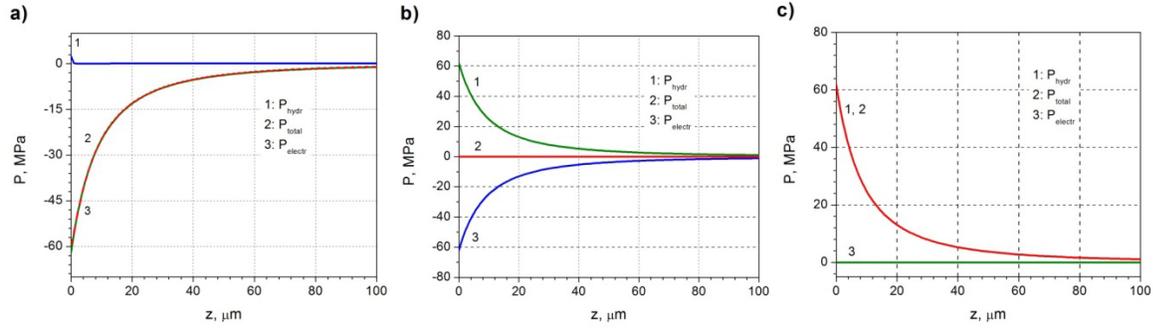

Figure 2 Pressure distributions in water in the high voltage electrode region at rise edge (a), plateau (b) and trailing edge of the voltage pulse.

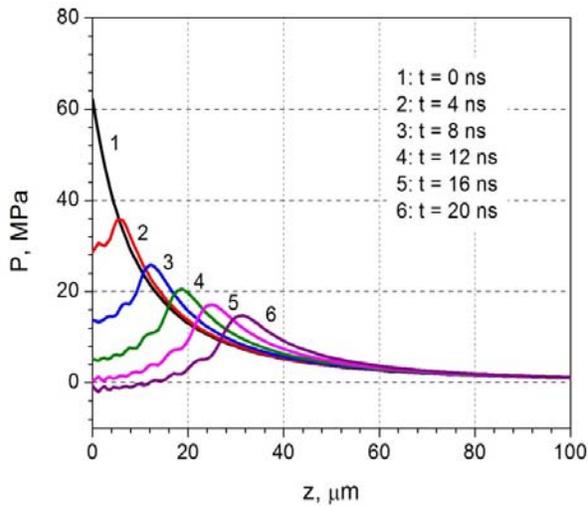

Figure 3 Longitudinal distributions of the hydrostatic pressure along the symmetry axis ($r=0$, $z$) at different time moments after the voltage pulse. Speed of wave is ~1.5 km/s.

Figure 4 shows the dependence of the total pressure in liquid in the vicinity of high voltage electrode at the point of maximum electric field on the rising edge for the case of water ($\varepsilon = 80$), ethanol ($\varepsilon = 25$) and their mixture ($\varepsilon = 55$). Due to fast rise time of the electric field, total pressure for the case of ethanol-water mixture and pure ethanol is negative, as in the case of pure water. However, absolute value of the maximum negative pressure is significantly lower compared to water, and is in a good agreement with the estimations made in [11]. It is also expected that no continuity disturbances (nanopores) occur in the case of pure ethanol (as the critical negative pressure is expected to be ~20 MPa).

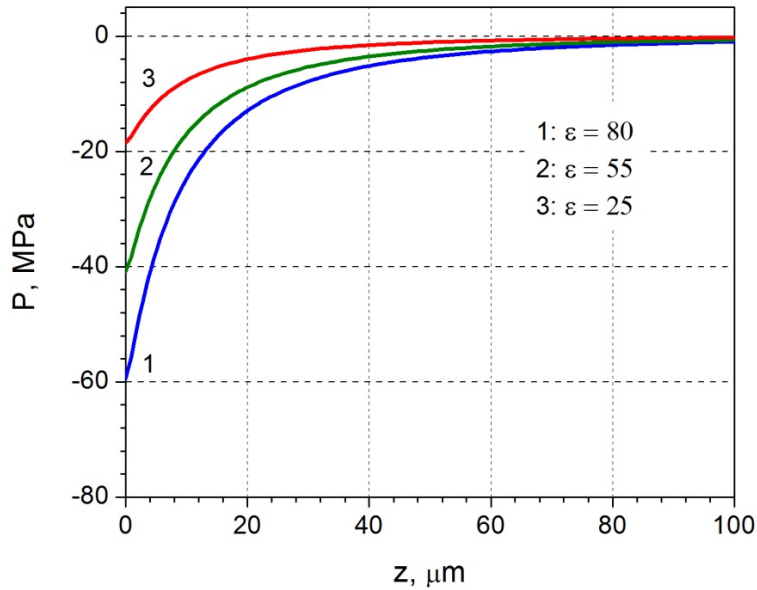

Figure 4 Total pressure in liquid in the vicinity of the high voltage electrode at the time of maximum electric field (point β on Figure 1, a) for water, ethanol and their 1-to-1 volume mixture.

## 3. Results and discussion

The experiments were conducted using 10 ns duration pulses with rise time of 3 ns and fall time of 4 ns. In all experiments, schlieren images were obtained using ICCD camera with 10 ns exposure time and 20 accumulations at the frequency of 1 Hz. Applied voltage in all cases was set to 11.2 kV which is below the breakdown voltage (i.e., the discharge was never ignited – the delay time between voltage application to the high voltage electrode and the discharge ignition is so small, that it was impossible to distinguish any liquid perturbations during the discharge phase). The results of the experiments conducted with pure water are shown on Figure 5. As it was predicted by the model, formation of a region with density perturbation in the vicinity of the high voltage electrode was observed. Immediately after the voltage pulse, a distinct shock wave formed, propagating symmetrically away from the electrode with the speed of $1.4 \pm 0.2$ km/s.

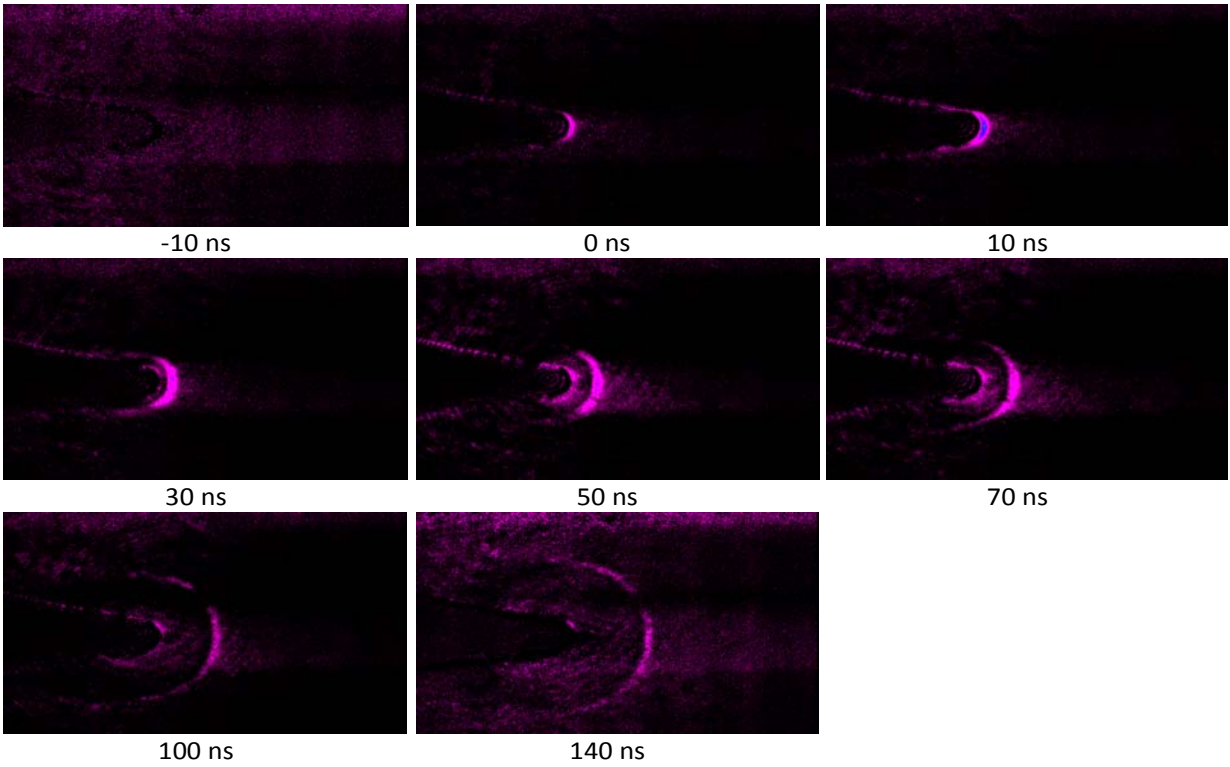

Figure 5 Schlieren images of the high voltage electrode region in water: 10 ns exposure, 20 accumulations. Image size 750x500 μm.  Shock wave propagation velocity is 1.4 ± 0.2 km/s.

In the case of water-ethanol mixture, which has lower dielectric permittivity than that of water, formation of a low-density region as well propagation of shock wave was also observed. In this case, propagation velocity of shock wave propagation was slightly higher – about 1.5 ± 0.3 km/s, and overall image and shock wave were more pronounced and contrast. As it was predicted by the model, no distinct region of density perturbation above noise was recorded in the case of pure ethanol (Figure 7).

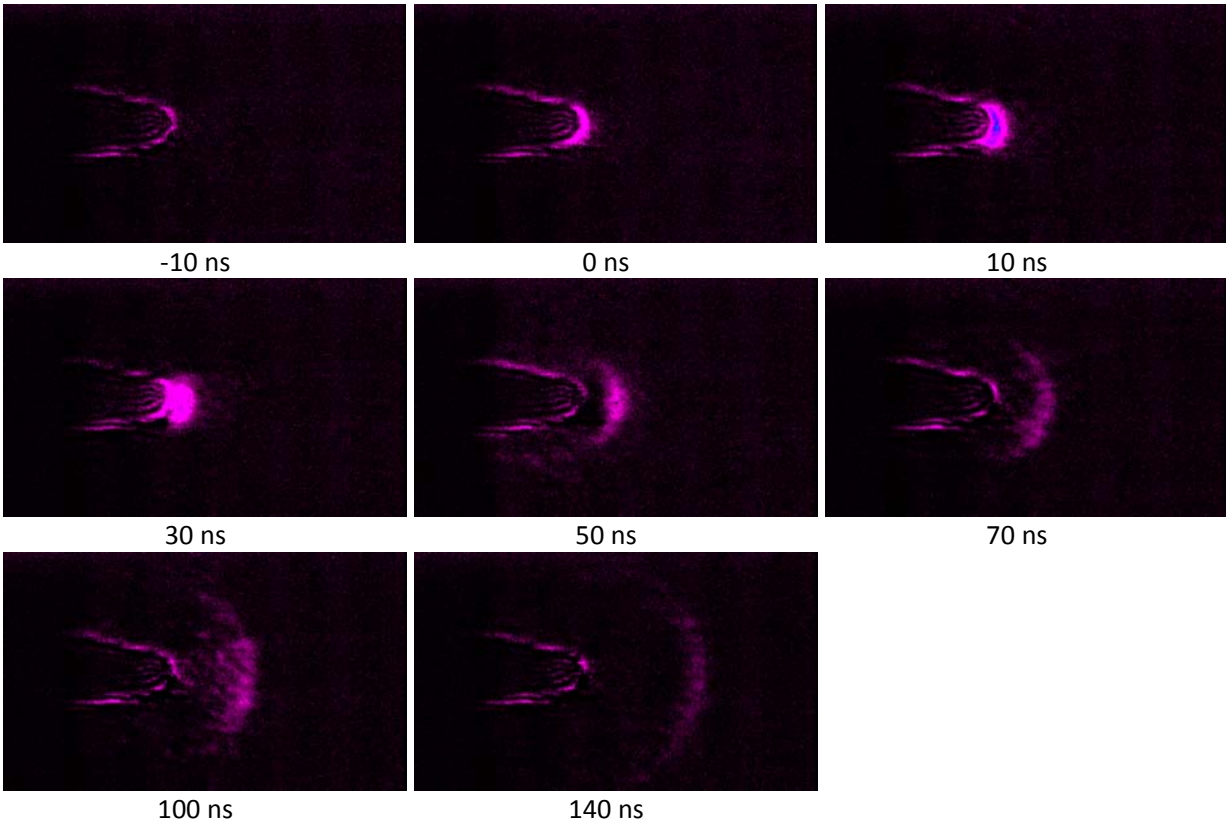

Figure 6 Schlieren images of the high voltage electrode region in ethanol-water mixture: 10 ns exposure, 20 accumulations. Image size 750x500 μm. Shock wave propagation velocity is 1.5 ± 0.3 km/s.

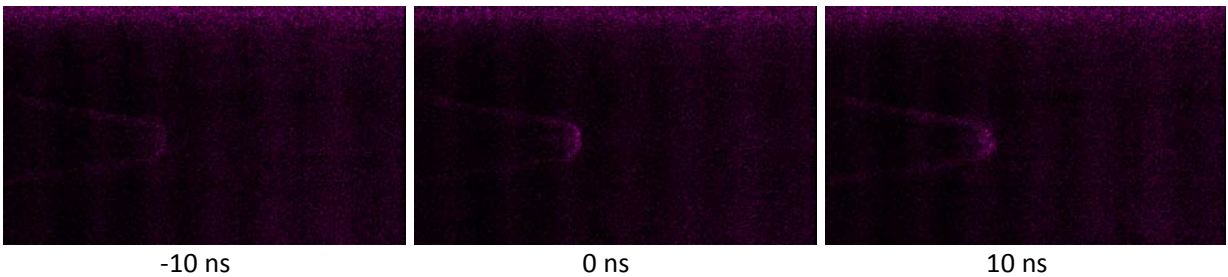

Figure 7 Schlieren images of the high voltage electrode region in ethanol: 10 ns exposure, 20 accumulations. Image size 750x500 μm.

The experimental results shown above are in a good agreement with the electrostriction model predictions. It is shown that application of strong inhomogeneous electric field with fast rise time leads to formation of a region saturated with nanopores – nanoscale irregularities in fluid structure due to negative pressure, - followed with formation of a shockwave propagating with the speed of sound. This electrostrictive effect is caused by rearrangement of the orientation of the elementary dipoles and occurs over a time much shorter than the characteristic time of the hydrodynamic processes. It follows that there is a possibility of formation of discharge in the liquids in time scales much shorter than the formation time for bubbles near the electrode. Thus, if the value of the applied electric field is high enough to create the minimum electron number that has to be generated for avalanche-to-streamer transition (this value corresponds to the well-known Meek's criterion, or $N_{e\,min} \sim 2 \times 10^8$). At the later stage, the discharge may develop similarly to the "long sparks" - by leader mechanism: once plasma channel is formed (if local electric field is sufficient for an avalanche-to-streamer transition), it grows due to high electric field of its charged tip – the conditions for electrostriction phenomenon are fulfilled in the vicinity of the channel's tip, providing ionization. Similarly to the leader mechanism, the channel grows as long as leader propagation is sustained by the constantly rising applied electric field. A significant portion of the applied voltage drops on the plasma channel with finite conductivity, and as a result, the field at the streamer head is not enough to maintain its propagation ("dark phase", see [7]).

## Acknowledgment

This work was supported by Defense Advanced Research Projects Agency (grant # DARPA-BAA-11-31).